\documentclass
[superscriptaddress,secnumarabic,amssymb,amsmath,nobibnotes,aps,prd,showkeys,nofootinbib,nopacsnumber,nopacs,onecolumn,12pt]{revtex4}%
\usepackage{graphicx}
\usepackage{epsf}
\usepackage{bm}
\usepackage{amsmath}
\usepackage{amsfonts}
\usepackage{amssymb}%
\providecommand{\U}[1]{\protect\rule{.1in}{.1in}}

\newcommand{\be}{\begin{equation}}
\newcommand{\ee}{\end{equation}}

\newcommand{\mincir}{\raise
-3.truept\hbox{\rlap{\hbox{$\sim$}}\raise4.truept\hbox{$<$}\ }}
\newcommand{\magcir}{\raise
-3.truept\hbox{\rlap{\hbox{$\sim$}}\raise4.truept\hbox{$>$}\ }}

\begin{document}
\title{Anisotropic spacetimes in $f(T,B)$ theory III: LRS Bianchi III Universe}
\author{Genly Leon}
\email{genly.leon@ucn.cl}
\affiliation{Departamento de Matem\'{a}ticas, Universidad Cat\'{o}lica del Norte, Avda.
Angamos 0610, Casilla 1280 Antofagasta, Chile.}
\affiliation{Institute of Systems Science, Durban University of Technology, PO Box 1334,
Durban 4000, South Africa}
\author{Andronikos Paliathanasis}
\email{anpaliat@phys.uoa.gr}
\affiliation{Institute of Systems Science, Durban University of Technology, PO Box 1334,
Durban 4000, South Africa}
\affiliation{Instituto de Ciencias F\'{\i}sicas y Matem\'{a}ticas, Universidad Austral de
Chile, Valdivia 5090000, Chile}

\begin{abstract}
We study the asymptotic dynamics of $f(T, B)$-theory in an anisotropic Bianchi
III background geometry. We show that an attractor always exists for the field
equations, which depends on a free parameter provided by the specific $f(T,
B)$ functional form. The attractor is an accelerated spatially flat FLRW or
non-accelerated LRS Bianchi III geometry. Consequently, the $f(T, B)$-theory
provides a spatially flat and isotropic accelerated Universe.

\end{abstract}
\date{\today}
\maketitle

\section{Introduction}

\label{sec1}

The family of spatially homogeneous Bianchi cosmologies includes an important
gravitational model, such as the Mixmaster Universe or the isotropic
Friedmann--Lema\^{\i}tre--Robertson--Walker (FLRW) spacetimes
\cite{mr,Mis69,mis1,mx1}. Nine anisotropic Bianchi models exist based on the
three-dimensional real Lie algebra classification. They act as isometries. In
these spacetimes, three-dimensional hypersurfaces are defined by the orbits of
three isometries. An important characteristic of the Bianchi models is that
the physical variables depend only on the time variable. The latter means that
the field equations are a system of ordinary differential equations
\cite{mx2,mx3}.

The FLRW spacetimes follow as the limit for some Bianchi models where the
anisotropy vanishes. Indeed, the flat, the open and the closed FLRW geometries
are related to the Bianchi I, the Bianchi III and the Bianchi IX spacetimes,
respectively \cite{WE}. In general, the Bianchi spacetimes are defined by
three scale factors \cite{mr} however, the locally rotational spacetimes (LRS)
admit an extra fourth isometry, and the LRS Bianchi line elements admit two
independent scale factors. It is interesting to mention that the LRS Bianchi
IX spacetime is related to the Kantowski-Sachs geometry \cite{ks1}.

We have devoted a series of papers to obtain conditions under which the
$f(T,B)$-model anisotropic model tends to the homogeneous and isotropic FRW
model. A related question is how the parameters and initial conditions of the
model influence the isotropization process. For example, it is well-known that
inflation is the most successful candidate to explain why the observable
Universe is currently homogeneous and isotropic with great precision. However,
the problem is not completely solved in the literature. That is, one usually
assumes from the beginning that the Universe is homogeneous and isotropic, as
given by the FLRW metrics, and then examines the evolution of the
perturbations, rather than starting with an arbitrary metric, and showing that
inflation does occur and that the Universe evolves towards homogeneity and
isotropy. The complete analysis is complex, even using numerical tools
\cite{Goldwirth:1989pr}. Thus, one should impose another assumption to extract
analytical information: consider anisotropic but homogeneous cosmologies. This
class of geometries \cite{Misner1973} exhibits very interesting cosmological
features, both in inflationary and postinflationary epochs
\cite{peebles1993principles}. Along these lines, isotropization is a crucial
question. Finally, the class of anisotropic geometries has recently gained
much interest due to anisotropic anomalies in the Cosmic Microwave Background
(CMB) and large-scale structure data, with strong evidence of a violation of
the Cosmological Principle in its isotropic aspect \cite{Fosalba:2020gls,
LeDelliou:2020kbm}. The Bianchi I spacetime reduces to the spatially flat FLRW
geometry, while the Bianchi III and the Kantowski-Sachs geometries reduce to
the open and closed FLRW geometries. Furthermore, Kantowski-Sachs geometry can
be naturally separated from Bianchi I and III since it gives a closed model,
Bianchi I is flat, and Bianchi III is open. The different geometries provide
different topologies.

This third work analyses the dynamics of modified teleparallel $f\left(
T,B\right)  $-theory in anisotropic spacetimes. We determine selection rules
in which initial conditions with anisotropy and curvature can lead to an
isotropic and accelerated spatially flat FLRW geometry within the $f\left(
T,B\right)  $-theory. In \cite{paper1} we performed a detailed analysis on the
dynamics for the Bianchi I Universe, while in \cite{paper2} we focused on the
Kantowski-Sachs geometry. Paper \cite{paper1} is the first part of a series of
studies on analyzing the $f(T,B)$-theory considering some anisotropic
spacetimes. The previous analysis of $f(T,B)$-gravity was reviewed, and the
global dynamics of a locally rotational Bianchi I background geometry were
investigated. A criterion for solving the homogeneity problem in the $f\left(
T,B\right)  $-theory was deduced. Finally, the integrability properties for
the field equations were investigated by applying the Painlev\'{e} analysis,
obtaining an analytic solution in terms of a right Painlev\'{e} expansion. On
the other hand, in \cite{paper2}, we construct a family of exact anisotropic
solutions while investigating the evolution of the field equations using
dynamical system analysis in $f\left(  T,B\right)  =T+F\left(  B\right)  $
theory of gravity. From the analysis, it follows that in this theory, with
initial conditions of Kantowski-Sachs geometry, the future attractor of the
Universe, can be a spatially flat spacetime which describes acceleration. The
de Sitter spacetime exists for a specific value of the free parameter.

In these two analyses, we found that future attractors exist where the
Universe is spatially flat and isotropic. We perform a similar analysis for
the LRS Bianchi III Universe in the following. For previous studies of Bianchi
III Universes in various gravitational theories we refer the reader to
\cite{b1,b2,b3,b4,b5,b6,b7,b8,b9,b10,b11,tl4} and references therein.\ The
application of teleparallelism in anisotropic geometries is presented in
detail in \cite{paper1}, and \cite{paper2}. Thus we continue with the plan of
the present paper.

In Section \ref{sec2} we briefly present the gravitational field equations of
$f\left(  T,B\right)  $-theory. In\ Section \ref{sec3} we focus on the
LRS\ Bianchi III Universe, and we derive the field equations. Some exact
solutions of special interest are studied in Section \ref{sec4}. Section
\ref{sec5} includes the main results of this analysis, where we present a
detailed study of the dynamics of the field equations. Finally, in\ Section
\ref{sec6} we discuss our results.

\section{$f\left(  T,B\right)  $ gravity}

\label{sec2}

In this section, we briefly discuss the field equations in the modified
teleparallel theory of our consideration; more details can be found in
\cite{revtel} or in the previous article of this series of studies
\cite{paper1}.

In $f\left(  T,B\right)  $-theory the gravitational Action Integral
is\ defined \cite{bh1}%
\begin{equation}
S_{f\left(  T,B\right)  }=\frac{1}{16\pi G}\int d^{4}xef\left(  T,B\right)
\label{cc.06}%
\end{equation}
where $T$ is the torsion scalar for the Weitzenb{\"{o}}ck connection
\cite{Weitzenb23} and $B=2e^{-1}\partial_{\nu}\left(  eT_{\rho}^{~\rho\nu
}\right)  $. The Ricciscalar $R~$and the torsion scalar $T$ are related as
$B=T+R$.

Variation of the Action Integral (\ref{cc.06}) for the vierbein fields lead to
the field equations%

\begin{align}
0  &  =ef_{,T}G_{a}^{\lambda}+\left[  \frac{1}{4}\left(  Tf_{,T}-f\right)
eh_{a}^{\lambda}+e(f_{,T})_{,\mu}S_{a}{}^{\mu\lambda}\right] \nonumber\\
&  +\left[  e(f_{,B})_{,\mu}S_{a}{}^{\mu\lambda}-\frac{1}{2}e\left(
h_{a}^{\sigma}\left(  f_{,B}\right)  _{;\sigma}^{~~~;\lambda}-h_{a}^{\lambda
}\left(  f_{,B}\right)  ^{;\mu\nu}g_{\mu\nu}\right)  +\frac{1}{4}%
eBh_{a}^{\lambda}f_{,B}\right]  . \label{cc.08}%
\end{align}

For the special case of~$f\left(  T,B\right)  =T+F\left(  B\right)  $ theory
which we will study in this work, we can define the scalar field $\phi$ and
the potential function as $\phi=F_{,B}$ and $V\left(  \phi\right)  =F-BF_{,B}$
such that the field equations can be written in the equivalent form
\begin{equation}
eG_{a}^{\lambda}+\left(  e\phi_{,\mu}S_{a}{}^{\mu\lambda}-\frac{1}{2}e\left(
h_{a}^{\sigma}\phi_{;\sigma}^{~~~;\lambda}-h_{a}^{\lambda}\phi^{;\mu\nu}%
g_{\mu\nu}\right)  +\frac{1}{4}eh_{a}^{\lambda}V\left(  \phi\right)  +\frac
{1}{4}eh_{a}^{\lambda}f\right)  =0
\end{equation}

We consider the background geometry to be that of LRS Bianchi III spacetime.

\section{LRS Bianchi III\ Universe}

\label{sec3}

The line element for the LRS\ Bianchi III Universe is
\begin{equation}
ds^{2}=-N^{2}\left(  t\right)  dt^{2}+e^{2\alpha\left(  t\right)  }\left(
e^{2\beta\left(  t\right)  }dx^{2}+e^{-\beta\left(  t\right)  }\left(
dy^{2}+\sinh^{2}\left(  y\right)  ~dz^{2}\right)  \right)  \label{ch.03}%
\end{equation}
where $N\left(  t\right)  $ is the lapse function,$~\alpha\left(  t\right)  $
is the scale factor for the three-dimensional hypersurface and $\beta\left(
t\right)  $ is the anisotropic parameter. For $\beta\left(  t\right)
\rightarrow0$, the line element (\ref{ch.03}) reduces to the closed FLRW geometry.

We assume the vierbein fields \cite{revtel}
\begin{align*}
e^{1}  &  =Ndt\\
e^{2}  &  =i~e^{\alpha+\beta}\cos z\sinh y~dx+e^{\alpha-\frac{\beta}{2}%
}\left(  \cosh y\cos z~dy-\sinh y\sin z~dz\right) \\
e^{3}  &  =i~e^{\alpha+\beta}\sin z~\sinh ydx+e^{\alpha-\frac{\beta}{2}%
}\left(  \cosh y\sin z~dy-\sinh y\cos z~dz\right) \\
e^{4}  &  =-e^{\alpha+\beta}\cosh y~dx-i~e^{\alpha-\frac{\beta}{2}}\sinh y~dy
\end{align*}
which provides%
\begin{equation}
T=\frac{1}{N^{2}}\left(  6\dot{\alpha}^{2}-\frac{3}{2}\dot{\beta}^{2}\right)
+2e^{-2\alpha+\beta}, \label{ch.04}%
\end{equation}
such that TEGR is recovered.

Moreover, the boundary term is calculated%
\begin{equation}
B=\frac{6}{N^{2}}\left(  \ddot{\alpha}-\dot{\alpha}\frac{\dot{N}}{N}%
+3\dot{\alpha}^{2}\right)  . \label{ch.05}%
\end{equation}

Hence, the modified gravitational field equations are%

\begin{equation}
0=6H^{2}-\frac{3}{2}\dot{\beta}^{2}-6H\dot{\phi}+V\left(  \phi\right)
-2e^{-2\alpha+\beta}, \label{ch.10}%
\end{equation}%
\begin{equation}
0=\dot{H}+3H^{2}+\frac{1}{6}V_{,\phi}~, \label{ch.11}%
\end{equation}%
\begin{equation}
0=\ddot{\beta}+3H\dot{\beta}+\frac{2}{3}e^{-2\alpha+\beta}~, \label{ch.12}%
\end{equation}
and%
\begin{equation}
0=\ddot{\phi}+3H^{2}+\frac{1}{2}V\left(  \phi\right)  +\frac{1}{3}V_{,\phi
}+\frac{3}{4}\dot{\beta}^{2}+\frac{1}{3}e^{-2\alpha+\beta}~. \label{ch.13}%
\end{equation}
Where without loss of generality we have selected $N=1$ and $H=\dot{a}$,

An important characteristic is the existence of a minisuperspace description
for the theory. Indeed, there exists the point-like Lagrangian
\begin{equation}
\mathcal{L}\left(  \alpha,\dot{\alpha},\beta,\dot{\beta},\phi,\dot{\phi
}\right)  =\frac{1}{N}\left(  e^{3\alpha}\left(  6\dot{\alpha}^{2}-\frac{3}%
{2}\dot{\beta}^{2}\right)  -6e^{3\alpha}\dot{\alpha}\dot{\phi}\right)
+Ne^{3\alpha}V\left(  \phi\right)  +2Ne^{\alpha+\beta}~, \label{ch.09}%
\end{equation}
which generates the field equations. The existence of this point-like
Lagrangian is essential because techniques from Analytic Mechanics can be
applied for the study of the field equations, such is the Noether symmetry
analysis for the determination of conservation laws or the quantization
process, such approaches are discussed in \cite{paper4}. \qquad

\section{Exact solutions}

\label{sec4}

Let us now investigate the existence of a power-law solution for the field
equations (\ref{ch.10})-(\ref{ch.13}). We assume that $\alpha=\alpha_{0}\ln t$
then, equation (\ref{ch.12}) ends with%
\begin{equation}
0=\ddot{\beta}+3\alpha_{0}t^{-1}\dot{\beta}+\frac{2}{3}e^{-\beta}%
e^{-2\alpha_{0}t}~.
\end{equation}
Hence, a closed-form solution of the latter equation is
\begin{equation}
\beta\left(  t\right)  =2\left(  \alpha_{0}-1\right)  \ln t~,~\alpha_{0}%
=\frac{2}{3}.
\end{equation}

Moreover, by replacing in the rest of the field equations we end with the
system%
\begin{equation}
V\left(  \phi\right)  =-4t^{-1}\dot{\phi}~,
\end{equation}
and%
\begin{equation}
t\ddot{\phi}-2\dot{\phi}=0.
\end{equation}

For the scalar field we find $\phi\left(  t\right)  =\frac{\phi_{1}}{3}%
t^{3}+\phi_{1}$, while for the scalar field potential it follows%
\begin{equation}
V\left(  \phi\right)  =-4~\phi_{1}^{\frac{2}{3}}\left(  3\left(  \phi-\phi
_{0}\right)  \right)  ^{\frac{1}{3}}.
\end{equation}

We proceed with the analysis of the dynamics.

\section{Asymptotic dynamics}

\label{sec5}

We define the dimensionless variables%
\begin{equation}
\Sigma=\frac{\dot{\beta}}{2H}~,~x=\frac{\dot{\phi}}{H}~,~y=\frac{V\left(
\phi\right)  }{H^{2}}~,~\Omega_{R}=\frac{e^{-2\alpha+\beta}}{3H^{2}}%
~,~\lambda=-\frac{V_{,\phi}}{V}\label{ch.14}%
\end{equation}
where as in the previous studies we assume $V\left(  \phi\right)
=V_{0}e^{-\lambda\phi}$, such that $\lambda=const$. The selection of the
exponential potential function is two-fold. In terms of dynamics, for such
potential function, the dimension of the dynamical system is reduced by one;
however this can provide the stationary points and for other potential
functions in the limit where $\lambda=const$, see for instance the discussion
in \cite{karp1}. Additionally, the exponential potential is of special
interest in terms of an isotropic universe. In previous studies,
\cite{Paliathanasis:2017efk,Paliathanasis:2017flf}; it was found that such
potential is cosmological viable and can explain the main epochs of the
cosmological evolution. Finally, the exponential potential has been found to
provide integrable cosmological equations in the case of isotropic and
spatially flat universe \cite{karp1}.

Thus, in the new variables $\left(  \Sigma,x,y,\eta\right)  $ the field
equations are written as the following system of algebraic-differential
equations%
\begin{equation}
\frac{d\Sigma}{d\tau}=-\lambda y\Sigma-\Omega_{R}~,\label{ch.15}%
\end{equation}%
\begin{equation}
\frac{dx}{d\tau}=3\left(  \Sigma^{2}-1\right)  +\left(  2\lambda-3\right)
y+x\left(  3-\lambda y\right)  -\Omega_{R}~,\label{ch.16}%
\end{equation}%
\begin{equation}
\frac{dy}{d\tau}=-y\left(  \lambda\left(  x+2y\right)  -6\right)
~,\label{ch.17}%
\end{equation}
and%
\begin{equation}
\frac{d\Omega_{R}}{d\tau}=2\left(  2-\lambda y+\Sigma\right)  \Omega
_{R}~,\label{ch.18}%
\end{equation}
with algebraic equation
\begin{equation}
1-x-y-\Sigma^{2}-\Omega_{R}=0,\label{ch.19}%
\end{equation}
and $d\tau=Hdt$.

Furthermore,~the deceleration parameter $q=-1-\frac{\dot{H}}{H^{2}}$ is
calculated%
\begin{equation}
q\left(  \Sigma,x,y,\eta\right)  =2-\lambda y\text{.} \label{ch.20}%
\end{equation}

We determine the stationary points for the dynamical system (\ref{ch.15}%
)-(\ref{ch.19}). Because of the constraint equation (\ref{ch.19}), the
dimension of the dynamical system can be reduced by one, and without loss of
generality, we select to replace $y=1-x-\Sigma^{2}-\Omega_{R}$ in the field
equations and we end with the system (\ref{ch.15}), (\ref{ch.16}) and
(\ref{ch.18}). For the asymptotic solution to describe a real solution, it
follows that $\Omega_{R}\geq0$. The rest of the variables are not a constraint.

\subsection{Analysis at the finite regime}

We replace $y$ from (\ref{ch.19}) in equations (\ref{ch.15}), (\ref{ch.16})
and (\ref{ch.18}) and we end with the system%
\begin{equation}
\frac{d\Sigma}{d\tau}=-\lambda y\Sigma-\Omega_{R}~,
\end{equation}%
\begin{equation}
\frac{dx}{d\tau}=3\left(  \Sigma^{2}-1\right)  +\left(  2\lambda-3\right)
\left(  1-x-\Sigma^{2}-\Omega_{R}\right)  +x\left(  3-\lambda\left(
1-x-\Sigma^{2}-\Omega_{R}\right)  \right)  -\Omega_{R}~,
\end{equation}%
\begin{equation}
\frac{d\Omega_{R}}{d\tau}=2\left(  2-\lambda\left(  1-x-\Sigma^{2}-\Omega
_{R}\right)  +\Sigma\right)  \Omega_{R}~,
\end{equation}
where now the deceleration parameter becomes%
\begin{equation}
q\left(  \Sigma,x,\eta\right)  =2-\lambda\left(  1-x-\Sigma^{2}-\Omega
_{R}\right)  .
\end{equation}

The stationary points $A=\left(  \Sigma\left(  A\right)  ,x\left(  A\right)
,\Omega_{R}\left(  A\right)  \right)  ~$and the corresponding eigenvalues for
the dynamical system are%
\[
A_{1}=\left(  \Sigma,\left(  1-\Sigma^{2}\right)  ,0\right)  ~,
\]
with eigenvalues
\begin{equation}
e_{1}\left(  A_{1}\right)  =0~,~e_{2}\left(  A_{1}\right)  =2\left(
\Sigma+2\right)  ~,~e_{3}\left(  A_{1}\right)  =\left(  6+\lambda\left(
\Sigma^{2}-1\right)  \right)  .
\end{equation}

The asymptotic solutions described by the family of points $A_{1}$ are that of
anisotropic Bianchi I spacetimes. Because for $\Omega_{R}=0,$the dynamical
system is reduced to that of Bianchi I, the analysis presented in
\cite{paper1} is valid. From there, we infer that the anisotropic solutions of
$A_{1}$ are always unstable.%
\[
A_{2}=\left(  0,2-\frac{6}{\lambda},0\right)  ~,
\]
with eigenvalues
\begin{equation}
e_{1}\left(  A_{2}\right)  =\left(  \lambda-6\right)  ~,~e_{2}\left(
A_{2}\right)  =\left(  \lambda-6\right)  ~,~e_{3}\left(  A_{2}\right)
=2\left(  \lambda-4\right)  .
\end{equation}

The stationary point $A_{2}$ describes a spatially flat FLRW geometry with
acceleration when $\lambda<4$, while in the particular case in which
$\lambda=3$ the de Sitter spacetime is recovered. We note that at this point,
the kinetic part of the scalar field and the potential term contributes to the
cosmological fluid. Moreover, the point is a sink for $\lambda<4$.
\[
A_{3}=\left(  \frac{4-\lambda}{1+2\lambda},\frac{6\left(  \lambda-1\right)
}{\lambda\left(  1+2\lambda\right)  },\frac{3\left(  \lambda-4\right)  \left(
\lambda+2\right)  }{\left(  1+2\lambda\right)  ^{2}}\right)  ~
\]
with corresponding eigenvalues
\begin{equation}
e_{1}\left(  A_{3}\right)  =-\frac{3\left(  2+\lambda\right)  }{1+2\lambda
}~,~e_{2,3}\left(  A_{3}\right)  =\frac{-3\lambda-6\pm i\sqrt{3\left(
2+\lambda\right)  \left(  \lambda\left(  16\lambda-59\right)  -38\right)  }%
}{2\left(  1+2\lambda\right)  }~.
\end{equation}
The point is physically accepted when $\lambda\geq4$ and $\lambda\leq-2$. The
deceleration parameter is $q\left(  A_{3}\right)  =\frac{\lambda-4}%
{2\lambda+1}$ which means that $q\left(  A_{3}\right)  \geq0$ when the point
exists. Thus, there is no acceleration. Finally from the eigenvalues we infer
that for $\lambda>4$ the asymptotic solution is stable, that is, $A_{3}$ is a
sink, while for $\lambda<-2$, point $A_{3}$ is a saddle point.

\begin{figure}[th]
\centering\includegraphics[width=1\textwidth]{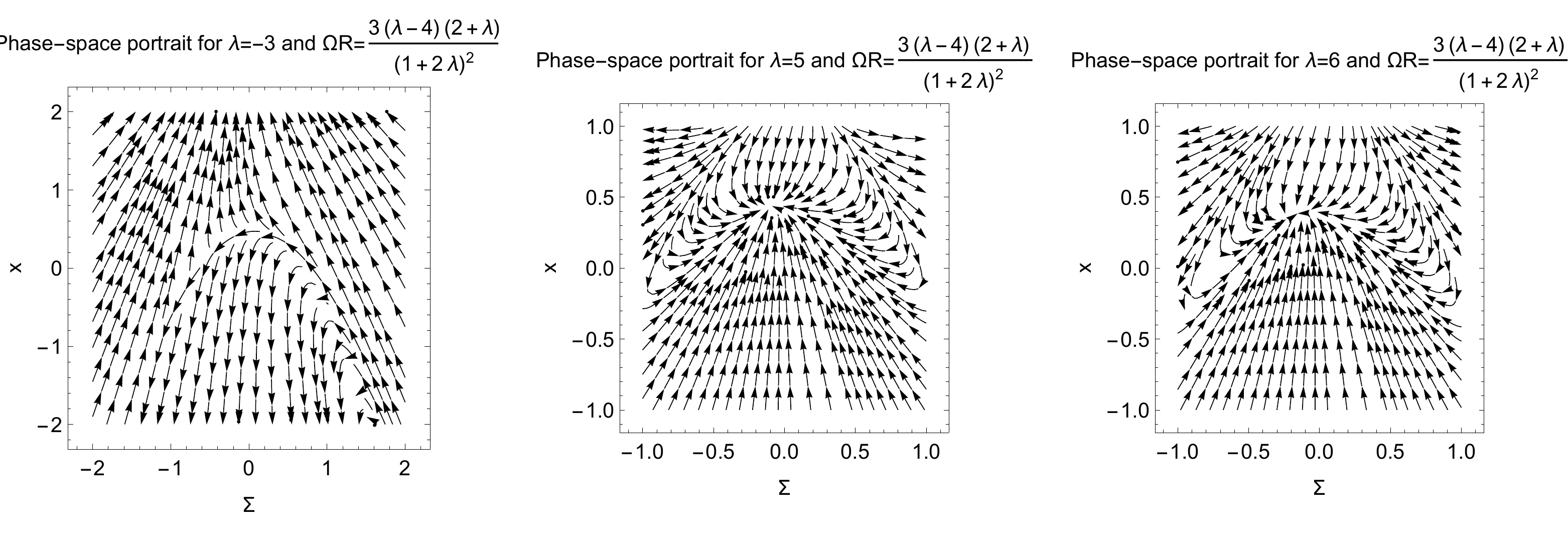}\caption{Phase-space
portrait for the dynamical system on the-dimensional surface $\left(
\Sigma,x\right)  $ for $\Omega_{R}=\frac{3\left(  \lambda-4\right)  \left(
\lambda+2\right)  }{\left(  1+2\lambda\right)  ^{2}}$ from where we observe
that the Kantoski-Sacks solution described by $A_{3}$ is always a saddle
point.}%
\label{b30}%
\end{figure}

In Fig. \ref{b30} we present phase-space portraits for the dynamical system on
the surface $\Omega_{R}=\frac{3\left(  \lambda-4\right)  \left(
\lambda+2\right)  }{\left(  1+2\lambda\right)  ^{2}}$, where it is clear that
for $\lambda>4$ the stationary point $A_{3}~$is an attractor.

Table \ref{tab1} summarizes the present analysis of the finite regime.%

\begin{table}[tbp] \centering
\caption{Stationary points at the finite regime}%
\begin{tabular}
[c]{ccccc}\hline\hline
\textbf{Point} & \textbf{Existence} & \textbf{Spacetime} & $\mathbf{q<0}$ &
\textbf{Stable?}\\\hline
$A_{1}$ & Always & Bianchi I & No & No\\
$A_{2}$ & $\lambda\neq0$ & FLRW (Flat) & $\lambda<4$ & $\lambda<4$\\
$A_{3}$ & $\lambda\geq4$ , $\lambda\leq-2$ & Bianchi III & No & $\lambda
>4$\\\hline\hline
\end{tabular}
\label{tab1}%
\end{table}%

\subsection{Analysis at the infinity}

We define the Poincar\'{e} variables%

\begin{equation}
x=\frac{\rho}{\sqrt{1-\rho^{2}}}\cos\Theta~,~\Sigma=\frac{\rho}{\sqrt
{1-\rho^{2}}}\sin\Theta\cos\Psi~, \label{sd.09}%
\end{equation}

\begin{equation}
\Omega_{R}=\frac{\rho^{2}}{1-\rho^{2}}\sin^{2}\Theta\sin^{2}\Psi
~,~d\sigma=\sqrt{1-\rho^{2}}d\tau~,
\end{equation}
with $\rho\in\left[  0,1\right]  $,~$\Theta\in\left[  0,\pi\right]  $ and
$\Psi\in\left[  0,\pi\right]  $

Therefore, the field equations read%
\begin{align}
4\frac{d\rho}{d\sigma}  &  =-4\rho^{4}\cos(\Theta)\left(  (\lambda
-2)\cos(2\Theta)+2\sin^{2}(\Theta)\cos(2\Psi)-2\lambda+8\right) \nonumber\\
&  +4\rho^{2}\cos(\Theta)\left(  (\lambda-2)\cos(2\Theta)+2\sin^{2}%
(\Theta)\cos(2\Psi)-4\lambda+14\right) \nonumber\\
&  -2\sqrt{1-\rho^{2}}\rho\left(  (2\lambda-5)\cos(2\Theta)+2\sin^{2}%
(\Theta)\cos(2\Psi)+4\lambda-7\right) \nonumber\\
&  -\frac{4\sin(\Theta)\left(  \rho\left(  \sqrt{1-\rho^{2}}\cos
(\Theta)(-2\lambda+\cos(2\Psi)+5)+(\lambda-2)\rho\cos(2\Theta)+2\rho\sin
^{2}(\Theta)\cos(2\Psi)\right)  \right)  }{\rho}\nonumber\\
&  -\frac{4\sin(\Theta)\left(  \lambda\left(  2-3\rho^{2}\right)  +8\rho
^{2}-6\right)  }{\rho}+8(\lambda-3)\cos(\Theta)\nonumber\\
&  +2\sqrt{1-\rho^{2}}\rho^{3}\left(  (\lambda-5)\cos(2\Theta)+2\sin
^{2}(\Theta)\cos(2\Psi)+5\lambda-7\right)  ~,
\end{align}%
\begin{align}
-\frac{2\rho}{\sin\left(  \Theta\right)  }\frac{d\Theta}{d\sigma}  &
=\rho^{2}(-(-2(\lambda-2)\cos(2\Theta)+\cos(2(\Theta-\Psi))+\cos(2(\Theta
+\Psi))\nonumber\\
&  +6\lambda-2\cos(2\Psi)-16))+2\sqrt{1-\rho^{2}}\rho\cos(\Theta
)(-2\lambda+\cos(2\Psi)+5)+4(\lambda-3)~,
\end{align}%
\begin{equation}
\frac{d\Psi}{d\sigma}=\sin(\Psi)\left(  \rho\sin(\Theta)+2\sqrt{1-\rho^{2}%
}\cos(\Psi)\right)  ~.
\end{equation}

Infinity is reached when $\rho=1$, thus the stationary points at the infinity
are defined on the two-dimensional surface $\left(  \Theta,\Psi\right)  $. The
points are%
\[
B^{1}=\left(  0,\Psi\right)  \text{ },\text{~}B^{2}=\left(  \pi,\Psi\right)
\]
for arbitrary $\lambda$, and
\[
D^{1}=\left(  \Theta,0\right)  \text{ },~D^{2}=\left(  \Theta,\pi\right)
\]
when $\lambda=3$. \ Similarly to the study presented in \cite{paper2} for the
Kantowski-Sachs Universe, the stationary points at the infinity, describe
isotropic spatially flat FLRW universes for arbitrary $\lambda$, and Bianchi I
geometry for $\lambda=3$. The eigenvalues for the stationary points are
\[
e_{1}\left(  B_{1}\right)  =0~,~e_{2}\left(  B_{1}\right)  =0~,~e_{3}\left(
B_{1}\right)  =-2\lambda~,
\]%
\[
e_{1}\left(  B_{2}\right)  =0~,~e_{2}\left(  B_{2}\right)  =0~,~e_{3}\left(
B_{2}\right)  =2\lambda~.
\]

\begin{figure}[th]
\centering\includegraphics[width=1\textwidth]{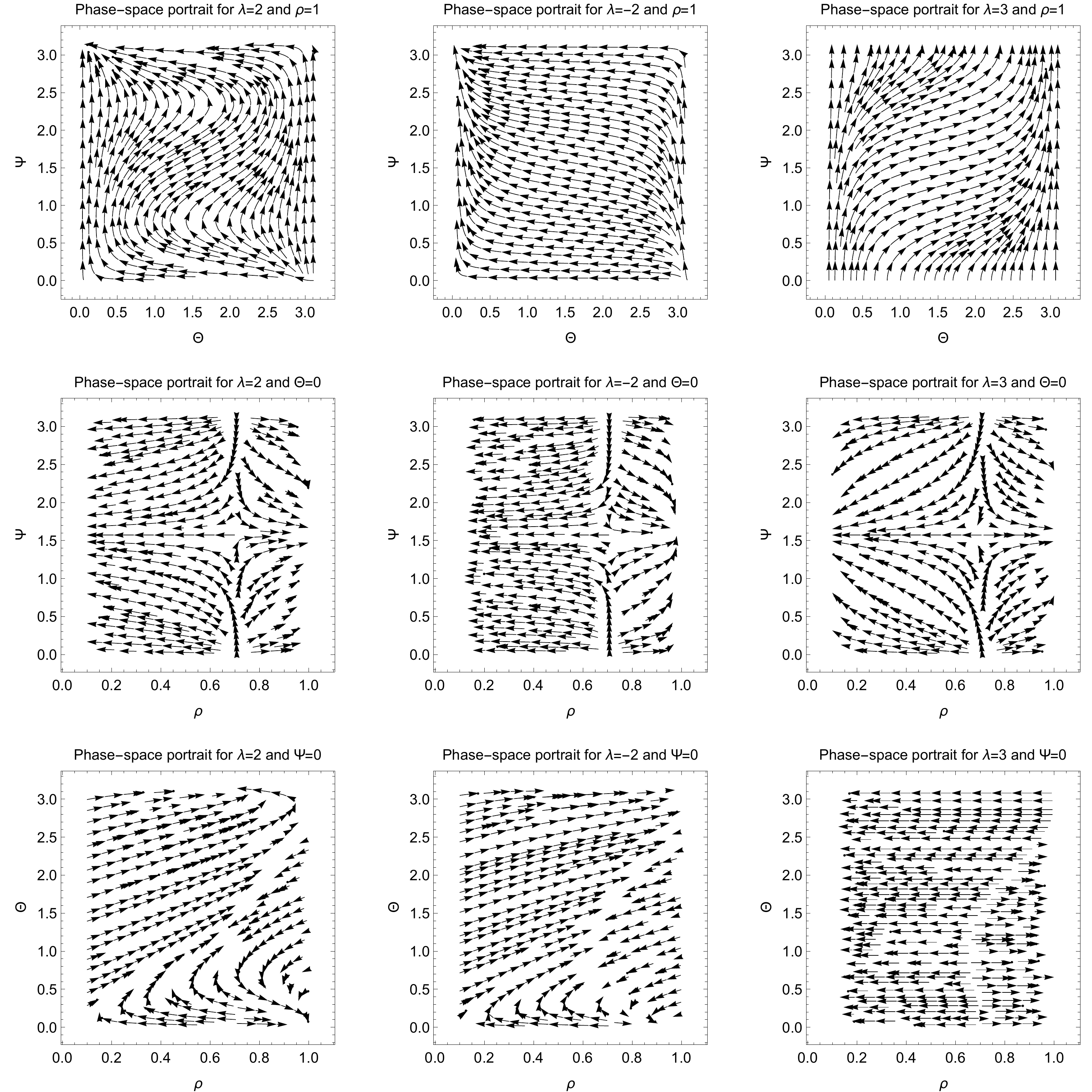}\caption{Phase-space
portrait for the three-dimensionless dynamical system on the Poincar\'{e}
variables. The plot clearly shows that the stationary points at the infinity
always describe unstable solutions.}%
\label{bp1}%
\end{figure}

We investigate the stability properties of the stationary points by using
numerical results. In Fig. \ref{bp1} we present two-dimensional phase-space
portraits for the dynamical system at the Poincare variables for various
values of parameters $\lambda$. From the evolution of the trajectories, it is
straightforward to conclude that the stationary points at the infinity always
describe unstable solutions. 

\section{Conclusions}

\label{sec6}

We performed a detailed analysis of the global dynamics of $f\left(
T,B\right)  $-theory for an LRS Bianchi\ III spacetime. Specifically, we
consider the $f\left(  T,B\right)  =T+F\left(  B\right)  $ theory where
$F\left(  B\right)  $ function can be seen that introduces small deviations
from the TEGR. The field equations can be written in the equivalent of a
second-order theory with a scalar field for this specific theory. This work is
part of our analysis of modified teleparallelism in anisotropic background geometries.

We determined the stationary points corresponding to asymptotic solutions for
the field equations. The stability properties were investigated such that to
construct the complete cosmological history. For the $F\left(  B\right)
=-\frac{1}{\lambda}B\ln B$, we found that for $\lambda<4$, the final attractor
of the field equations is a spatially flat and accelerated FLRW Universe,
where the de Sitter Universe is recovered for $\lambda=3$. However, for
$\lambda>4$ the future attractor is an anisotropic Universe described by the
Bianchi III geometry. These are the two unique attractors for the field
equations. The value $\lambda<4$ agrees with the previous studies on the
\ Bianchi I and Kantowski-Sachs geometries. For $\lambda<4$, the $f\left(
T,B\right)  $ can solve the flatness and isotropic problems by leading to an
accelerated Universe. Hence, the isotropization process of the Universe
depends only on the parameter $\lambda$.  We conclude that in $f\left(
T,B\right)  $ theory, the initial conditions which describe a homogeneous and
anisotropic open Universe can provide a spatially flat isotropic universe and
explain the inflationary epoch.

Last but not least, we mention that there are no attractors for the dynamical
the system at the infinity regime, and the trajectories of the field equations
have the origin at the finite and infinity regimes; thus, the attractors are
in the finite regime.

In \cite{paper4} we continue our study by applying the Noether symmetries for
the construction of conservation laws.

\textbf{Data Availability Statements:} Data sharing not applicable to this
article as no datasets were generated or analyzed during the current study.

\begin{acknowledgments}
The research of Genly Leon is funded by Vicerrector\'ia de Investigaci\'on y
Desarrollo Tecnol\'ogico at Universidad Cat\'olica del Norte.
\end{acknowledgments}


\bigskip

\begin{thebibliography}{99}                                                                                               %


\bibitem {mr}M.P. Ryan and L.C. Shepley, Homogeneous Relativistic Cosmologies,
Princeton University Press (1975)

\bibitem {Mis69}\ C.W. Misner, Astroph. J. 151, 431 (1968)

\bibitem {mis1}C.W. Misner, Phys. Rev. Lett. 22, 1071 (1969)

\bibitem {mx1}N. Cornish and J. Levin, Phys.\ Rev. D 55, 7486 (1997)

\bibitem {mx2}G.F.R. Ellis and M.A.H. MacCallum, Comm. Math. Phys. 12, 108 (1969)

\bibitem {mx3}M. Goliath and G.F.R. Ellis, Phys. Rev.\ D 60, 023502 (1999)

\bibitem {WE}J. Wainwright and G. F. R. Ellis, Dynamical Systems in Cosmology,
Cambridge University Press (1997)

\bibitem {ks1}R. Kantowski and R.K.\ Sachs, J. Math. Phys. 7, 443 (1966)

\bibitem {Goldwirth:1989pr}Goldwirth, D. S., and Piran, T. Inhomogeneity and
the Onset of Inflation.Phys. Rev. Lett. 64 (1990), 2852-2855

\bibitem {Misner1973}Misner, C. W., Thorne, K. S., and Wheeler, J.
A.Gravitation. 1973.

\bibitem {peebles1993principles}Peebles, P., and Peebles, P.Principles of
Physical Cosmology. Princeton Series in Physics. Princeton University Press, 1993.



\bibitem {Fosalba:2020gls}P.~Fosalba and E.~Gaztanaga,
doi:10.1093/mnras/stab1193 [arXiv:2011.00910 [astro-ph.CO]].




\bibitem {LeDelliou:2020kbm}M.~Le Delliou, M.~Deliyergiyev and A.~del Popolo,
Symmetry \textbf{12} (2020) no.10, 1741


\bibitem {paper1}A. Paliathanasis, Anisotropic spacetimes in $f(T,B)$ theory
I: LRS Bianchi I Universe (2022)

\bibitem {paper2}G.\ Leon and A. Paliathanasis, Anisotropic spacetimes in
$f(T,B)$ theory I: Kantowski-Sachs Universe (2022)

\bibitem {b1}S.S. Bayin and J.P. Krisch, J. Math. Phys. 27, 262 (1986)

\bibitem {b2}R. Tikekar and L.K. Patel, Gen.\ Rel.\ Grav. 24, 397 (1992)

\bibitem {b3}A. Feinstein and J. Ibanez, Class.Quant.Grav. 10, 93 (1993)

\bibitem {b4}S. Byland and D.\ Scialom, Phys. Rev. D 57, 6065 (1998)

\bibitem {b5}M. Tanimoto, V. Moncrief and K. Yasuno, Class. Quantum\ Grav. 20,
1879 (2003)

\bibitem {b6}U. Camci and Y. Kucukakca, Phys. Rev. D 76, 084023 (2007)

\bibitem {b7}A. Pradhan, S. Lata and H. Amirhashchi, Comm. Theor. Phys. 54,
950 (2010)

\bibitem {b8}S. Chandel and S.\ Ram, Indian J. Phys. 87, 1283 (2013)

\bibitem {b9}P.K. Sahoo, S.K. Sahu and A. Nath, Eur. Phys. J. Plus 131, 18 (2016)

\bibitem {b10}S.K. Banik, D.K. Banik and K. Bhuyan, Gen.\ Rel.\ Grav. 50, 24 (2018)

\bibitem {b11}G. Leon, E. Gonzalez, S. Lepe, C. Michea and A. Millano, EPJC
81, 414 (2021)

\bibitem {tl4}M. E. Rodrigues, A. V. Kpadonou, F. Rahaman, P. J. Oliveira and
M. J. S. Houndjo, Astroph. Sp. Sci. 357, 129 (2015)

\bibitem {Weitzenb23}R. Weitzenb\"{o}ck, Invarianten Theorie, Nordhoff,
Groningen (1923)

\bibitem {bh1}S. Bahamonte, C.G. Boehmer and M. Wight, Phys. Rev. D 92, 104042 (2015)

\bibitem {revtel}S. Bahamonte, K.F. Dialektopoulos, C. Escamilla-Rivera, V.
Gakis, M. Hendry, J.L. Said, J. Mifsud and E. Di Valentino, Teleparallel
Gravity: From Theory to Cosmology, [arXiv:2106.13793] (2021)

\bibitem {paper4}A. Paliathanasis, Anisotropic spacetimes in $f(T,B)$ theory
IV: Noether symmetry analysis (2022)



\bibitem {Paliathanasis:2017efk}A.~Paliathanasis,
Phys. Rev. D \textbf{95}, no.6, 064062 (2017)




\bibitem {Paliathanasis:2017flf}A.~Paliathanasis,
JCAP \textbf{08}, 027 (2017)


\bibitem {karp1}L. Karpathopoulos, S. Basilakos, G. Leon, A. Paliathanasis and
M. Tsamparlis, Gen. Rel. Grav. 50, 79 (2018)
\end{thebibliography}
\end{document}